\documentclass[sort&compress,5p]{elsarticle}

\usepackage{graphicx}
\usepackage{dcolumn}
\usepackage{bm}
\usepackage{amsmath}
\usepackage{multirow}
\newcommand{\astra}{{\sc Astra }}
\newcommand{\warp}{{\sc Warp }}

\usepackage[colorlinks=true]{hyperref}

\begin{document}

\title{Numerical Simulations of Early-Stage Dynamics of Electron Bunches Emitted from Plasmonic Photocathodes}

\author[niu]{A. Lueangaramwong\corref{cor1}}
\ead{anusorn@nicadd.niu.edu}
\author[niu]{D. Mihalcea}
\author[radia]{G. Andonian}
\author[niu,apc]{P. Piot}
\address[niu]{Department of Physics and Northern Illinois Center for Accelerator \& Detector Development, \\
Northern Illinois University, DeKalb, IL  60115, USA}
\address[radia]{Radiabeam Technologies LLC, Santa Monica, CA 90404, USA}
\address[apc]{Accelerator Physics Center, Fermi National Accelerator Laboratory, Batavia, IL  60510, USA}
\cortext[cor1]{Corresponding author}
\date{today}

\begin{abstract}
High-brightness electron sources are a key ingredient to the development of compact accelerator-based light sources. The electron sources are commonly based on (linear) photoemission process where a laser pulse with proper wavelength impinges on the surface of a metallic or semiconductor photocathode. Very recently, the use of plasmonic cathodes --  cathodes with a nano-patterned surface -- have demonstrated great enhancement in quantum efficiencies~\cite{li}.  Alternatively, this type of photocathodes could support the formation of structured beams composed of transversely-separated beamlets. In this paper we discuss numerical simulations of the early-stage beam dynamics of the emission process from plasmonic cathodes carried out using the \warp~\cite{warp1} framework. The model is used to investigate the properties of beams emitted from these photocathode and subsequently combined with particle-in-cell simulations to explore the imaging of cathode pattern after acceleration in a radiofrequency gun. 
\end{abstract}

%
\maketitle


\section{Introduction}
High-brightness electron sources play a crucial role in the development of compact accelerator-based light sources. A photoinjector typically combines a photocathode with a resonant radiofrequency (RF) cavity. The photocathodes often employed are metallic and require an ultraviolet laser pulse impinging on their surfaces. The development of photocathodes capable of attaining higher brightness or larger quantum efficiency is an active topic of research~\cite{Swanwick}. Most recently, the use of three-photon photoemission process was realized in photoinjector~\cite{pietro} and the associated electron-yield was shown to drastically improved by nano-engineering the cathode surface~\cite{li}. A sub-wavelength nano-patterned surface can increase the laser-pulse absorption thereby enhancing the overall quantum efficiency~\cite{nanolbnl}.

Following Ref.~\cite{li}, we consider throughout this paper a photocathode engineered to have a periodic array of nanoholes with Gaussian profile. The early-stage beam dynamics of the electron photoemitted from the nano-holes surface is investigated for different geometries. Our investigations are conducted in the weak-field regime (where quantum tunneling is insignificant) and implements a  multi-photon emission process in \warp~\cite{warp1,warp}. Our approach consists in first modeling a single nanohole  with proper electromagnetic boundary conditions and then replicating the generated macroparticle distribution to generate the beam distribution produced from the entire cathode surface (consisting of an array of nanoholes). 
\section{Electron-emission model}
All the simulations were carried out with an augmented version of \warp program which includes a modified emission model. The model is based on a ``piecewise" approach, which depending on the normalized vector potential of the incoming laser $a_0\equiv \frac{e E_0 \lambda}{2\pi mc^2}$ (where $E_0$ and $\lambda$ are respectively the laser peak E-field and wavelength, and $e$ and $mc^2$ the electronic charge and rest mass), applies a multi-photon or a field-emission model. In this paper we limit our studies to the weak field regime where the Keldish parameter $\gamma \equiv \sqrt{\Phi/(2U_p)} \gg 1$ (here $U_p$ and $\Phi$ are respectively the laser ponderomotive potential and material workfunction). In such a regime, our model follows the generalized Fowler-Dubridge's (FD) law and the photoemitted current density is 
\begin{eqnarray}~\label{eq:FD}
\pmb{j}_{FD}(\pmb x,t, \nu)= \sum_n {\pmb j}_n(\pmb x,t, \nu), 
\end{eqnarray}
where the partial current densities are given by $ \pmb{j}_n(\pmb x,t)= C_n |\pmb S(\pmb x,t)|^n, $ with $\pmb S(\pmb x,t)$ being the Poynting vector associated to the exciting laser evaluated  at the emitting surface and the constant $C_n$ compiles the material properties including reflection coefficient and multi-photon probability. The constants $C_n$ can be directly measured experimentally; see~\cite{pietro,li,nanolbnl}. Given Eq.~\ref{eq:FD}, the number of electrons emitted within a elementary surface area $d^2\pmb{A}$ at position $\pmb x$ at integration time step $t_i$, is 
\begin{eqnarray}
\delta N(\pmb x,t_i)=\frac{1}{|e|}\pmb{j}_{FD}(\pmb x,t_i).d^2\pmb{A}\times \delta t, 
\end{eqnarray}
where the elementary surface element is $d^2\pmb{A}= d^2A \hat{\pmb{n}}$, $ \delta t$ is the simulation time step, $\hat{\pmb{n}}$ is the normal vector to the elementary surface area which depends on the computational-domain mesh size. 
\begin{figure}[hhhhhh!!!!]
\begin{center}
\includegraphics[width=0.85\linewidth]{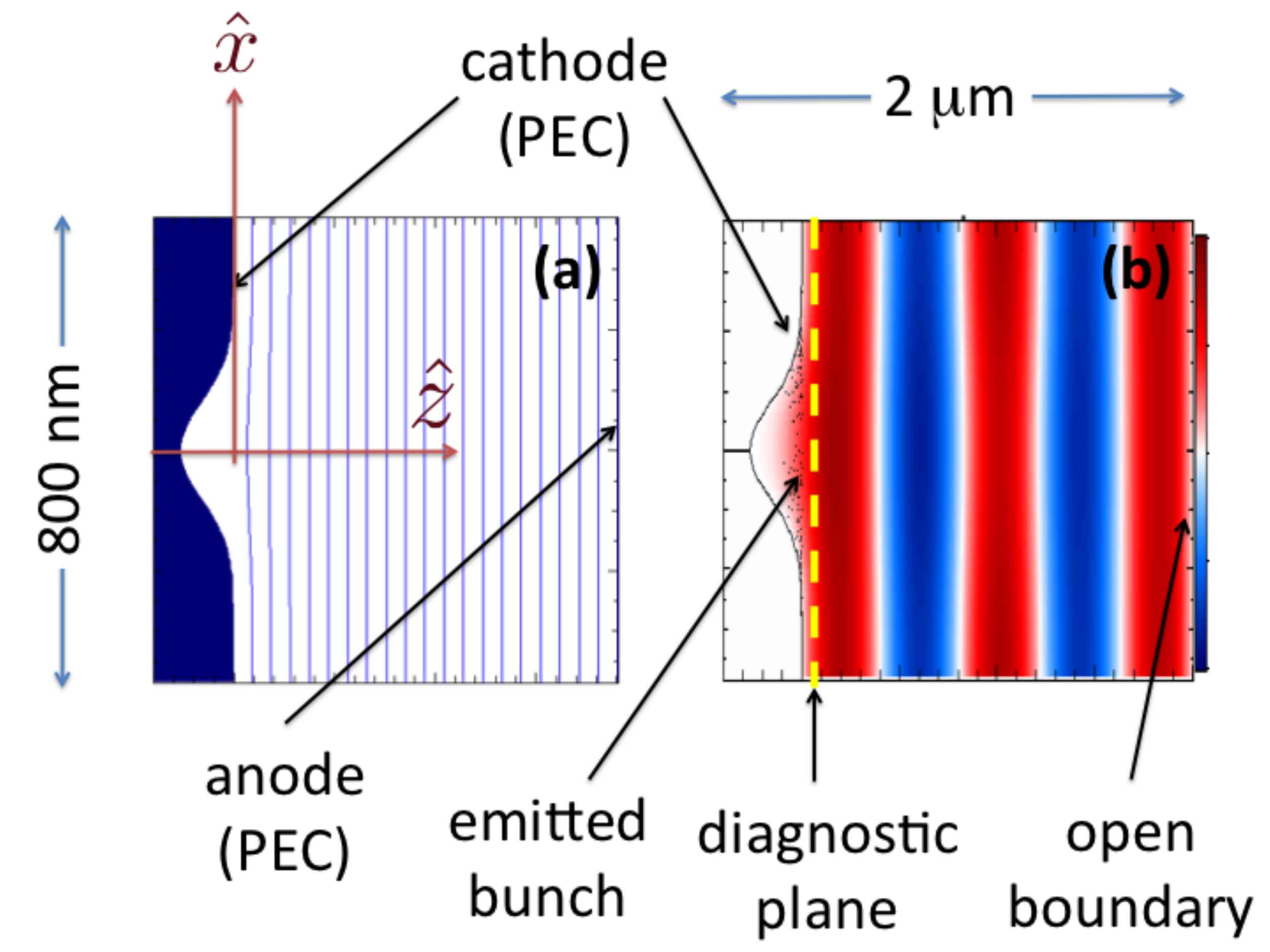}
\end{center}
\caption{\label{fig:geom} Geometries used to solve the electrostatic problem with associated potential lines (a) and configuration used with the electromagnetic solver showing the laser transverse field (colored contours) and emitted macroparticle (black dots) close to the cathode surface (b). The "PEC" stands for perfect-conductor }
\end{figure}

An example of implemented nanohole-cathode geometry appears in Fig.~\ref{fig:geom}(a-b). The Gaussian nanohole is described by the profile $z(r)=-d \exp [-r^2/(2w^2)]$, where $d$ and $w$ are respectively the hole depth and rms width and $r\equiv\sqrt{x^2+y^2}$ is the radial position with respect to the hole axis of symmetry. 

In a first step an electrostatic solver (ES) provides the electrostatic field configuration in the cathode vicinity. The cathode-anode potential difference $V$ is selected to mimic typical electric field sustained in radiofrequency (RF) guns ($E_0 \sim 100$~MV/m in S-band guns). An electromagnetic (EM) solver then simulates the propagation of a laser pulse launched at large $z$ and traveling in the $-\hat{\pmb {z}}$ direction; see Fig.~\ref{fig:geom}(b). The ES and EM fields are both used as external fields during the particle emission and dynamics. 

The EM solver also accounts for collective effects within the emitted particle bunch~\footnote{In this work a particle-in-cell (i.e. mean-field) approach is implemented. Future improvements will include Coulomb collisions}. 
The particle distribution is saved at plane at $z=100$~nm. Typical integration times are on the order of $\sim 300$~fs, i.e. 3 orders of magnitude smaller than the period of an S-band gun ($T \simeq 350$~ps for $f=2856$~MHz) thereby supporting our static-field approximation. Throughout this paper we take the laser duration to be 10~fs. 

\begin{figure}[b!!!!!!]
\begin{center}
\includegraphics[width=1.0\linewidth]{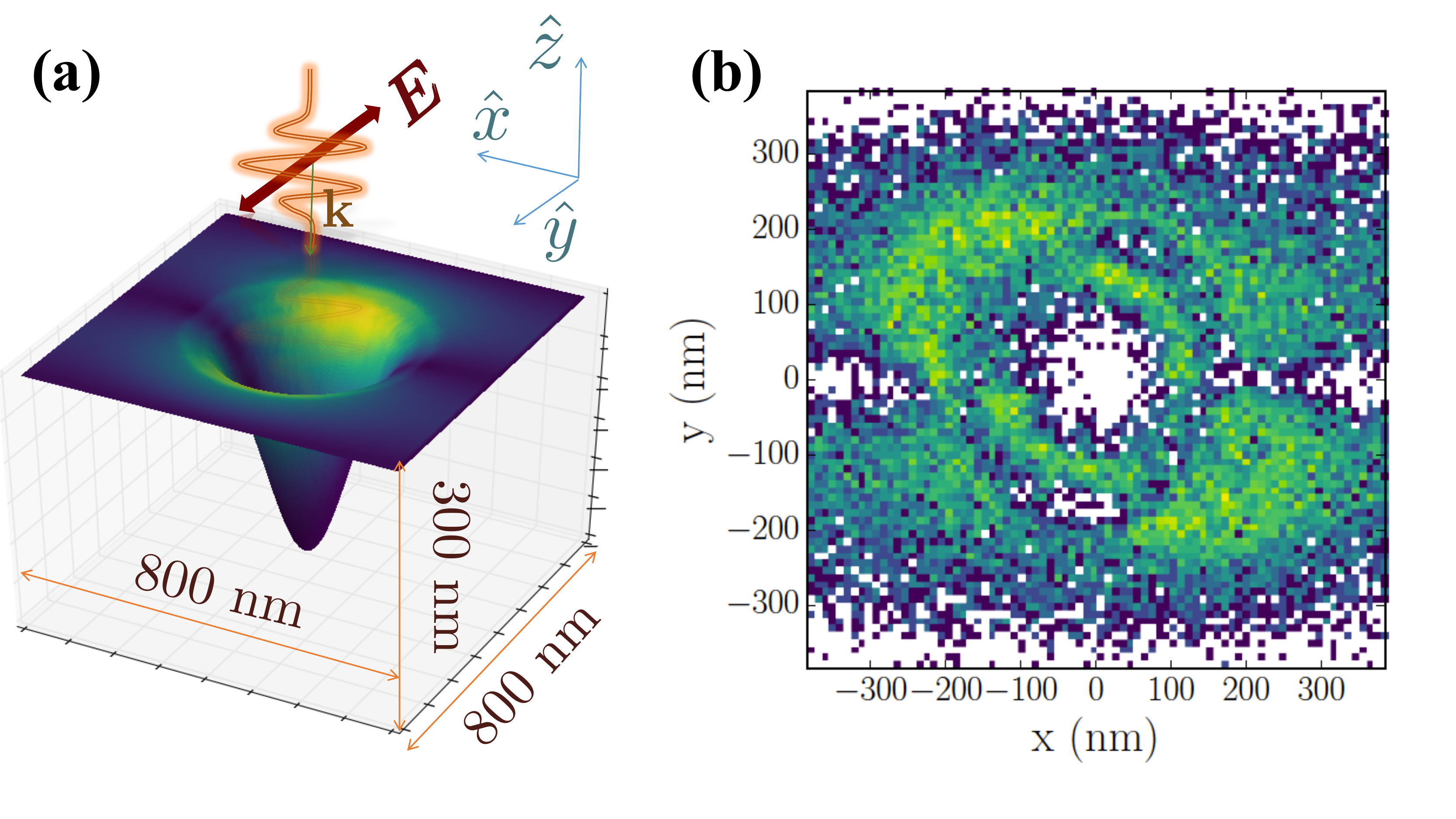}
\end{center}
\caption{\label{fig:fieldenhancement} Electromagnetic field (a) painted on the nanohole surface (the double arrow indicates the laser polarization, while $\pmb k$ is the laser wavevector). Emitted electrons recorded 100-nm away from the nanocathode surface (b). The nanohole depth $d$ and rms width $w$ are $d=264$ and $w=155$~nm, respectively. $a_0$ is fixed to $6\times 10^{-5}$}
\end{figure}

The EM solver indicates some E-field enhancement at the nanohole edges (at loci with largest local curvatures). The enhancement is  asymmetric with peak values occurring along the direction of the incoming-laser polarization; see Fig.~\ref{fig:fieldenhancement}. 

\section{Single-structure model}
\begin{figure}[hhhhhh!!!!!]
\includegraphics[width=0.49\linewidth]{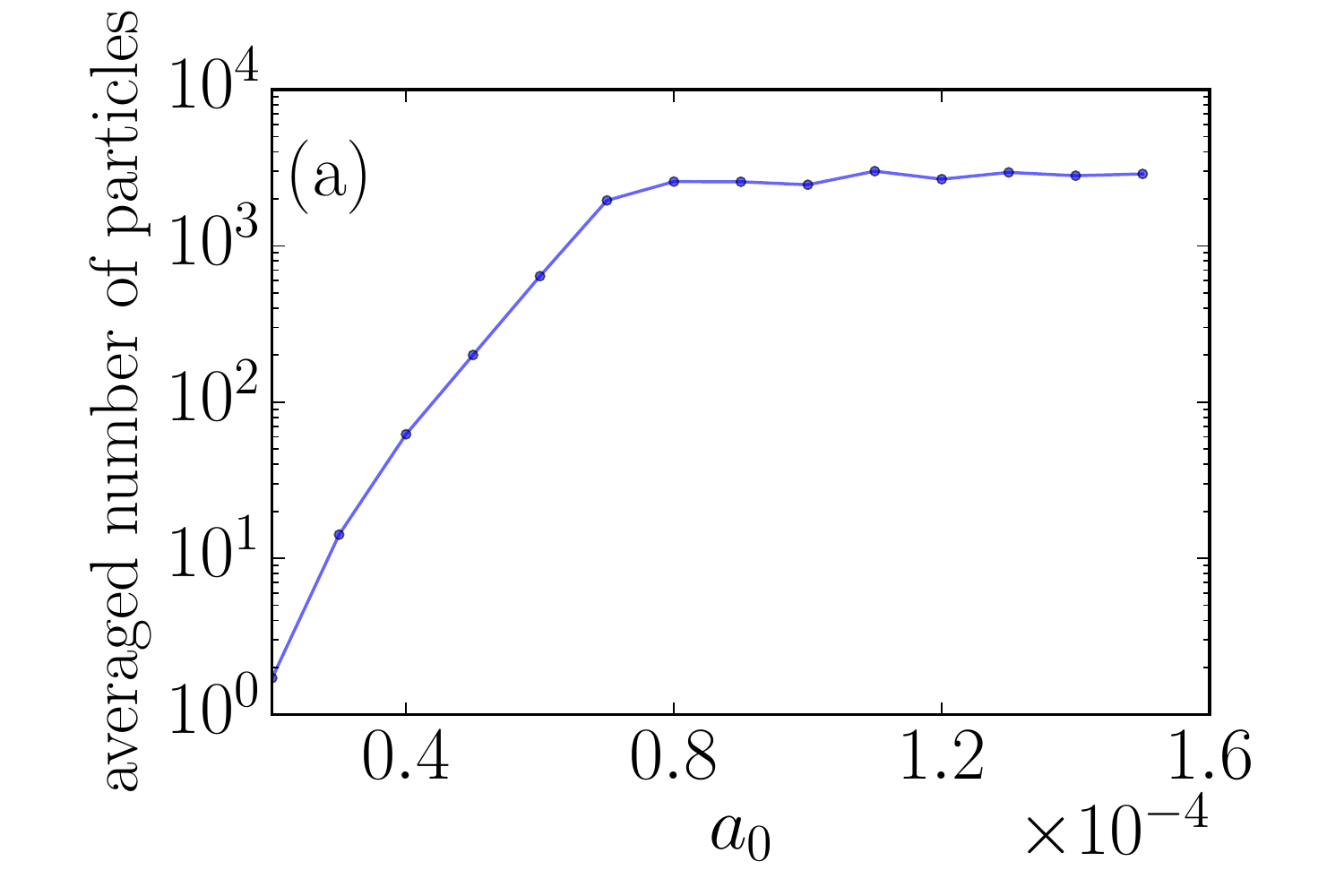}
\includegraphics[width=0.49\linewidth]{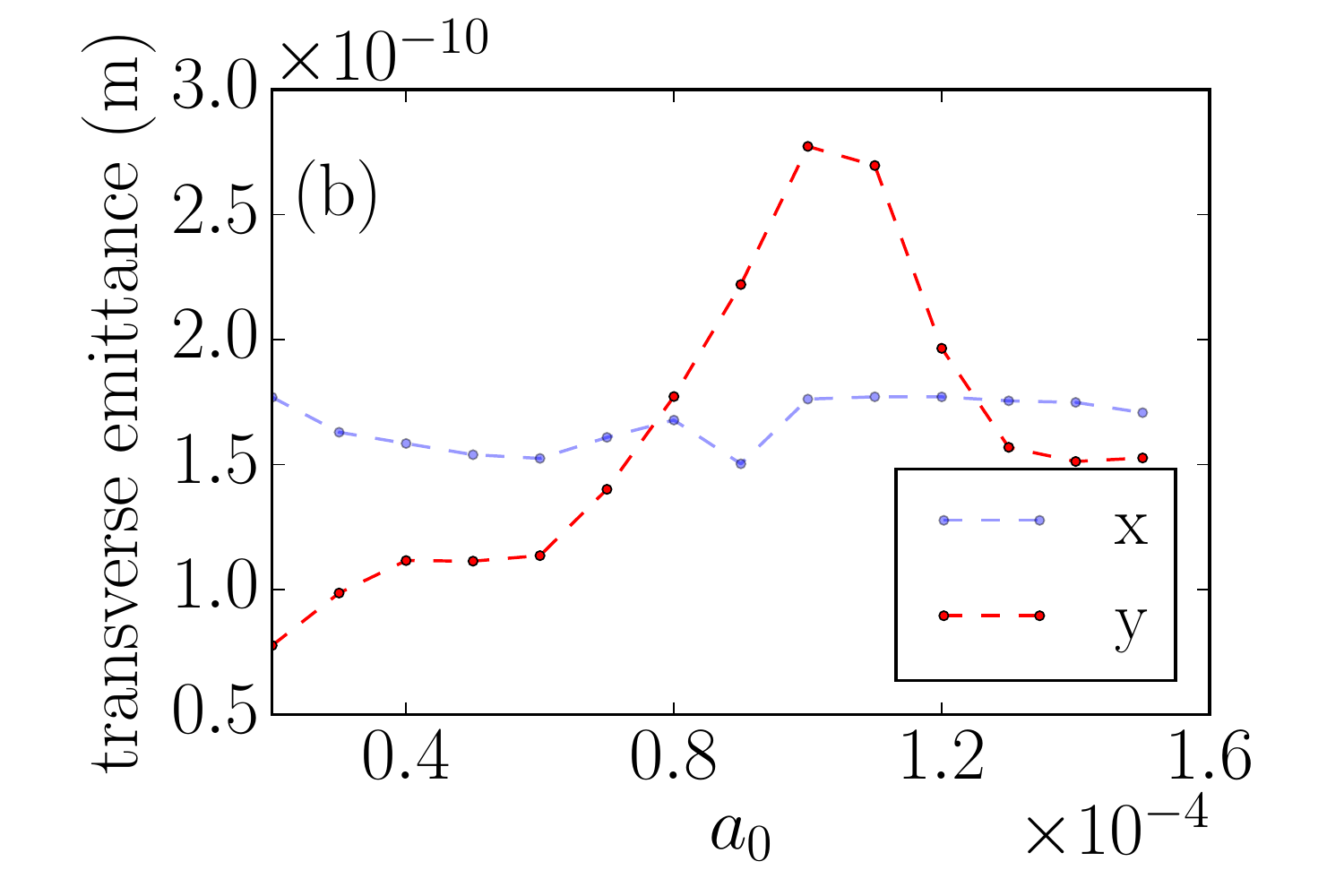}
\includegraphics[width=0.49\linewidth]{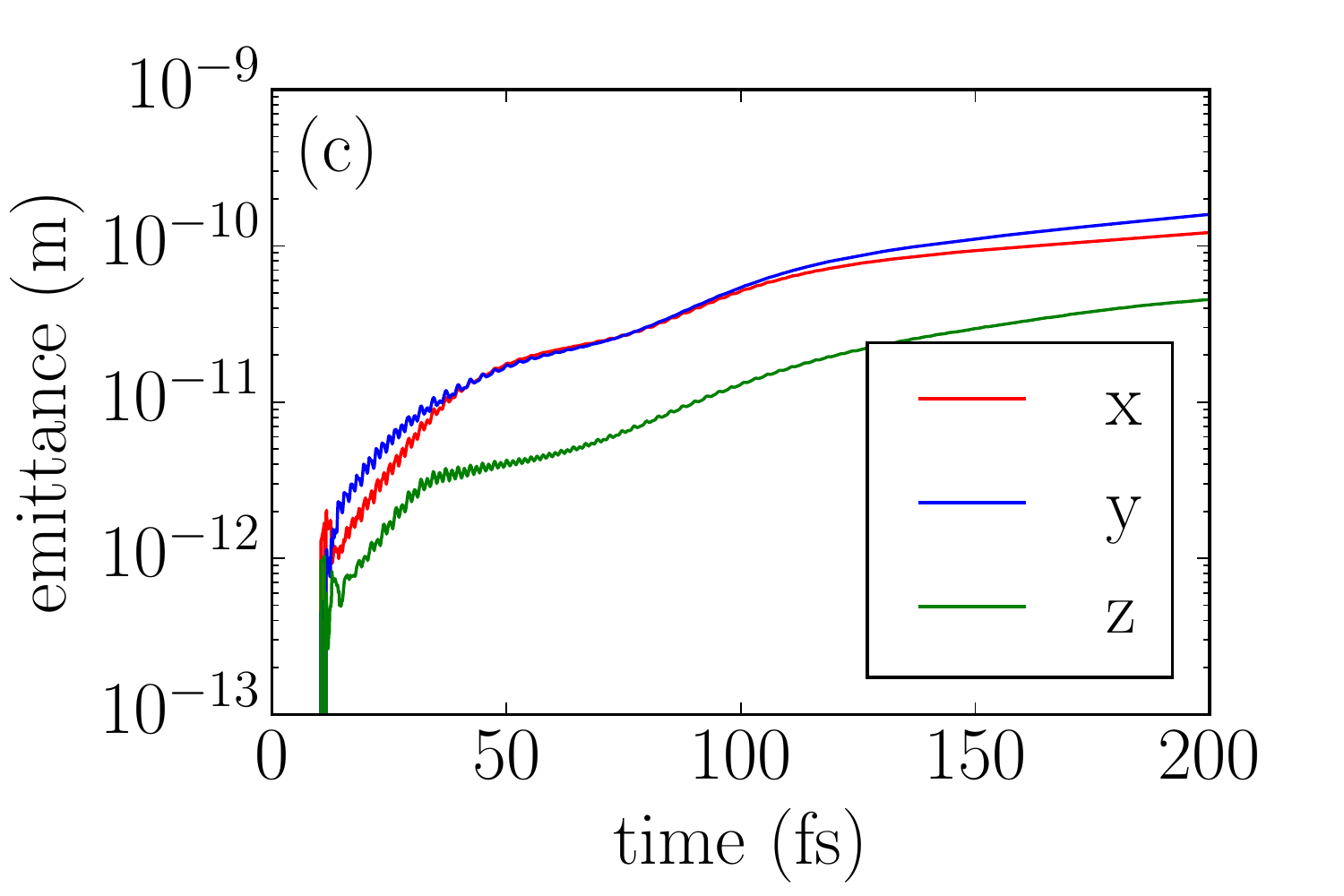}
\includegraphics[width=0.49\linewidth]{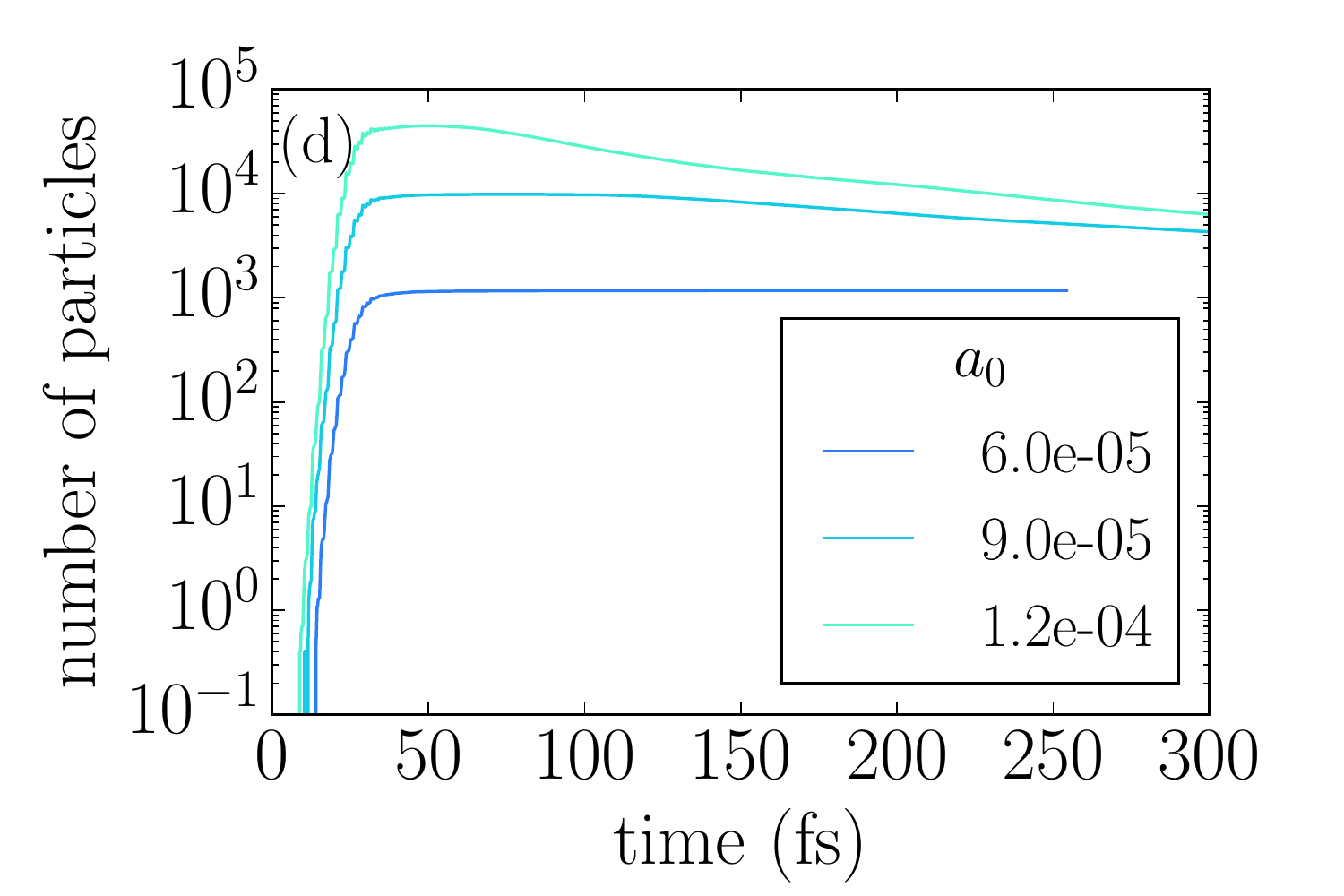}
\caption{\label{fig:evolution_emitt} Number of electron recorded 100-nm away from the nanocathode surface (a) and transverse-emittance  (b) as a function of $a_0$. Example of temporal evolution of the transverse emittance for $a_0 = 9\times10^{-5}$ (c) and number of emitted electrons (d). The nanohole geometry is fixed to $d=264$ and $w=155$~nm.}
\end{figure}

The beam-dynamics simulation of a cathode consisting of a periodic structure is first examined by considering a single-nanohole structure. Throughout this paper we limit our studies to three-photon emission process and consider  $C_3=10^3 $ A GW$^{-3}$ cm$^4$ consistent with Ref.~\cite{li}.  Figure~ \ref{fig:evolution_emitt} summarizes salient features of the beam-formation process. First the number of particles as function of $a_0$ saturates; see~Fig.~\ref{fig:evolution_emitt}(a). Such a saturation for $a_0\sim 6\times 10^{-5}$ is attributed to space-charge-limited emission. Further exploration of the charge evolution in time for three values of $a_0$ [see Fig.~\ref{fig:evolution_emitt}(b)] confirms the space charge emission limit (some early-emitted particles ``push" later-emitted particles back onto the cathode). For low value of $a_0$, normalized transverse and longitudinal emittances on the order of ${\cal O}(10^{-10})$~m are simulated. 
These emittance values are consistent with requirements needed for the compact light-source concept discussed in Ref.~\cite{graves,sun}. In the low $a_0$ regime, we consistently observed $\varepsilon_y<\varepsilon_x$. The reason for such an asymmetry comes from the field asymmetric enhancement along the laser polarization; see Fig.~\ref{fig:fieldenhancement}(a). Such a field configuration causes the emission of electron $|\pmb j_3 | \propto |\pmb E|^6$ to increase at the locations of maximum fields which yields an initially bi-modal distribution with narrower rms width along the $y$ direction with associated smaller emittance $\varepsilon_y$. As $a_0$ increases and space-charge comes at play, we find that the emission at this large field-amplitude are lowered and eventually result in a more uniform distribution resulting in  $\varepsilon_y\simeq \varepsilon_x$ for larger value of $a_0$.

\begin{figure}[hhhhhh!!!!!!]
	\includegraphics[width=0.49\linewidth]{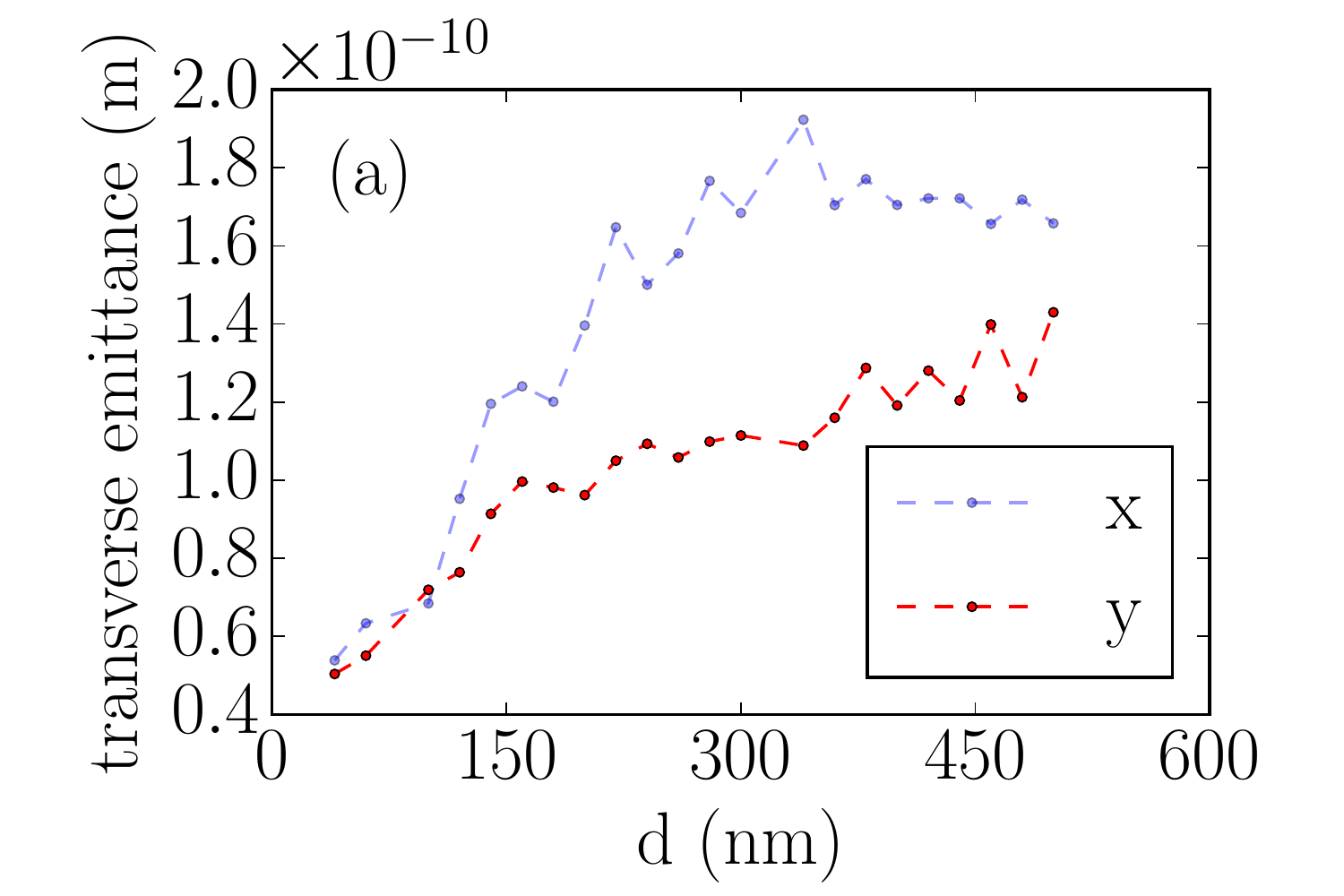}
	\includegraphics[width=0.49\linewidth]{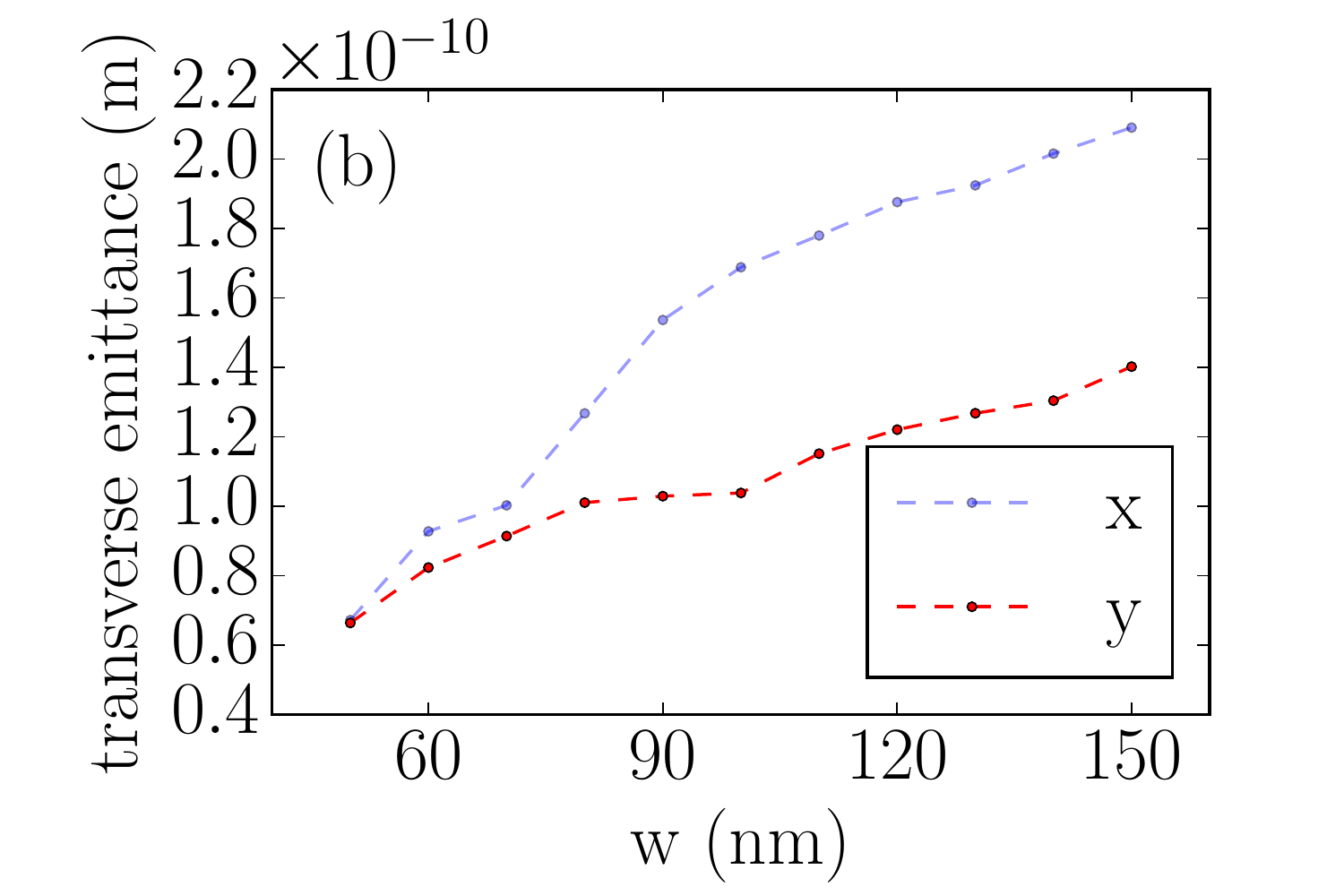}
	\includegraphics[width=0.49\linewidth]{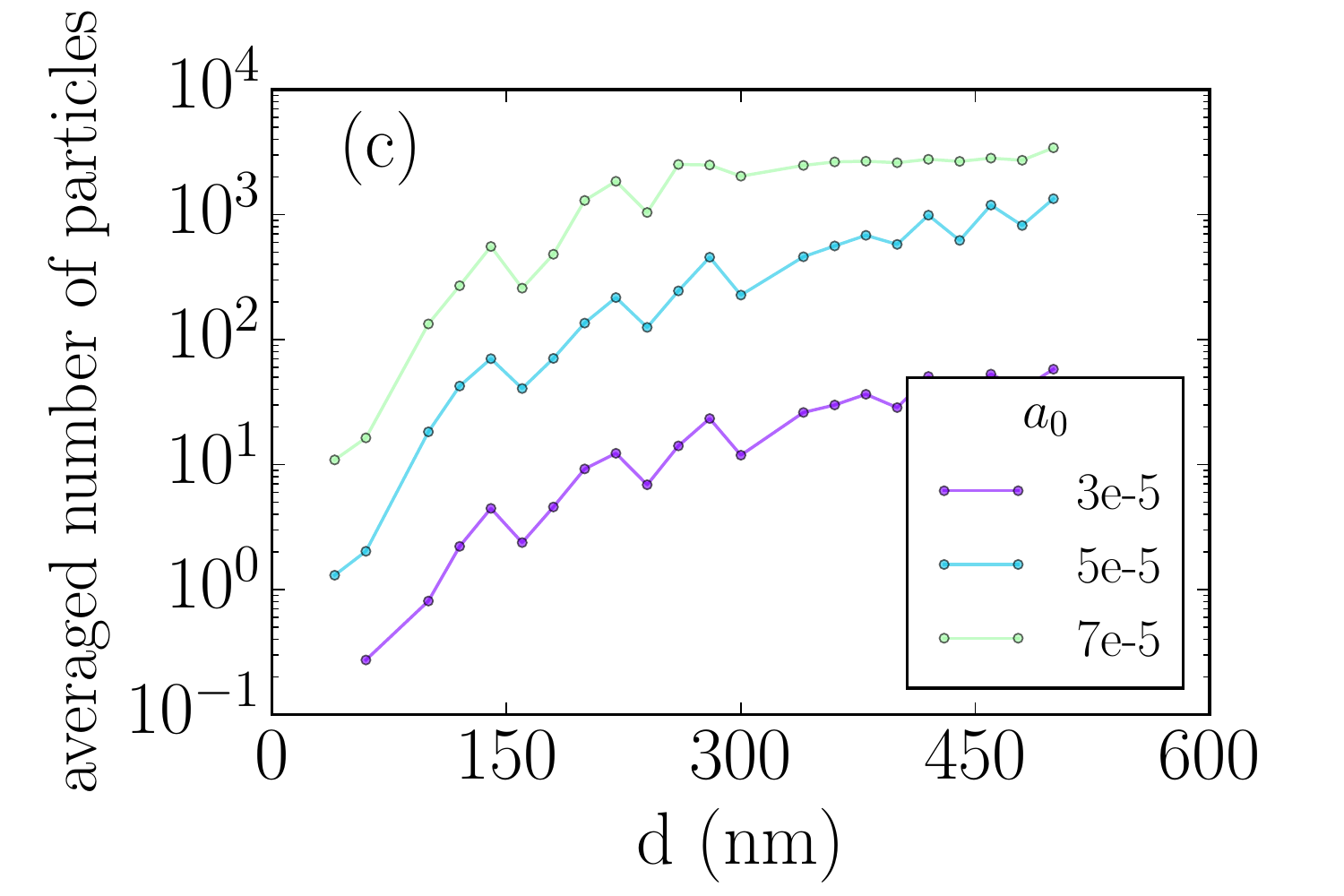}
	\includegraphics[width=0.49\linewidth]{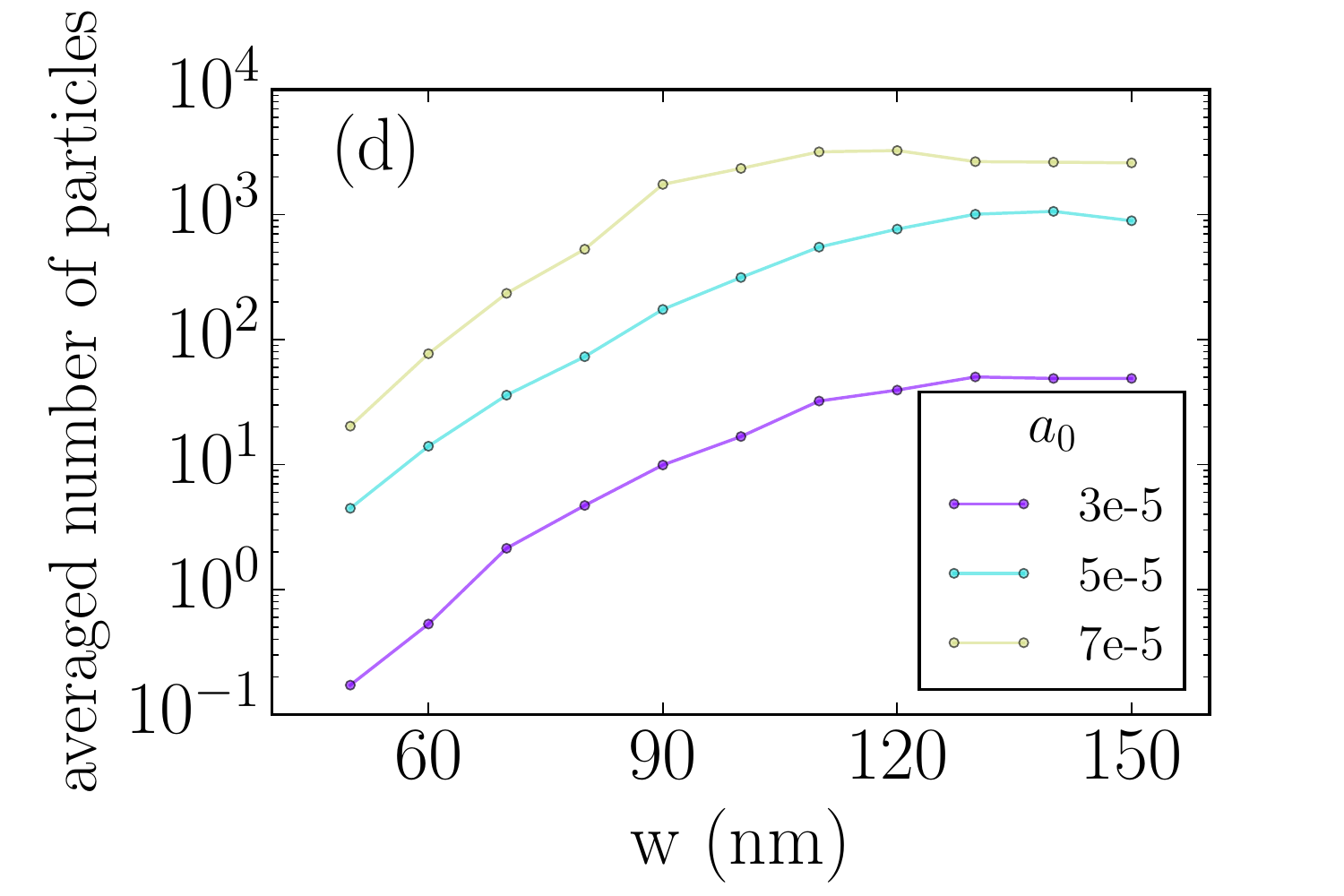}
	\includegraphics[width=0.49\linewidth]{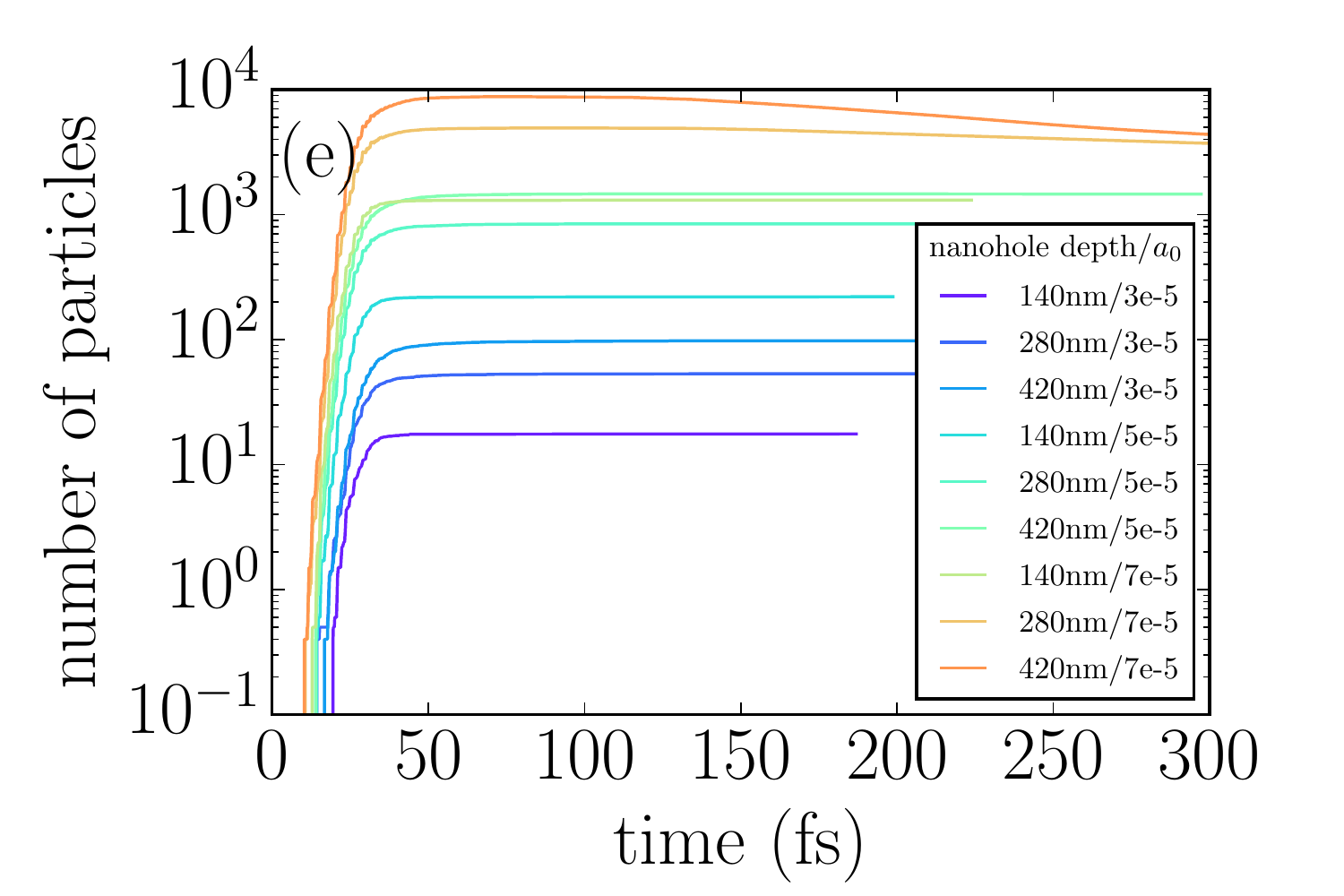}
	\includegraphics[width=0.49\linewidth]{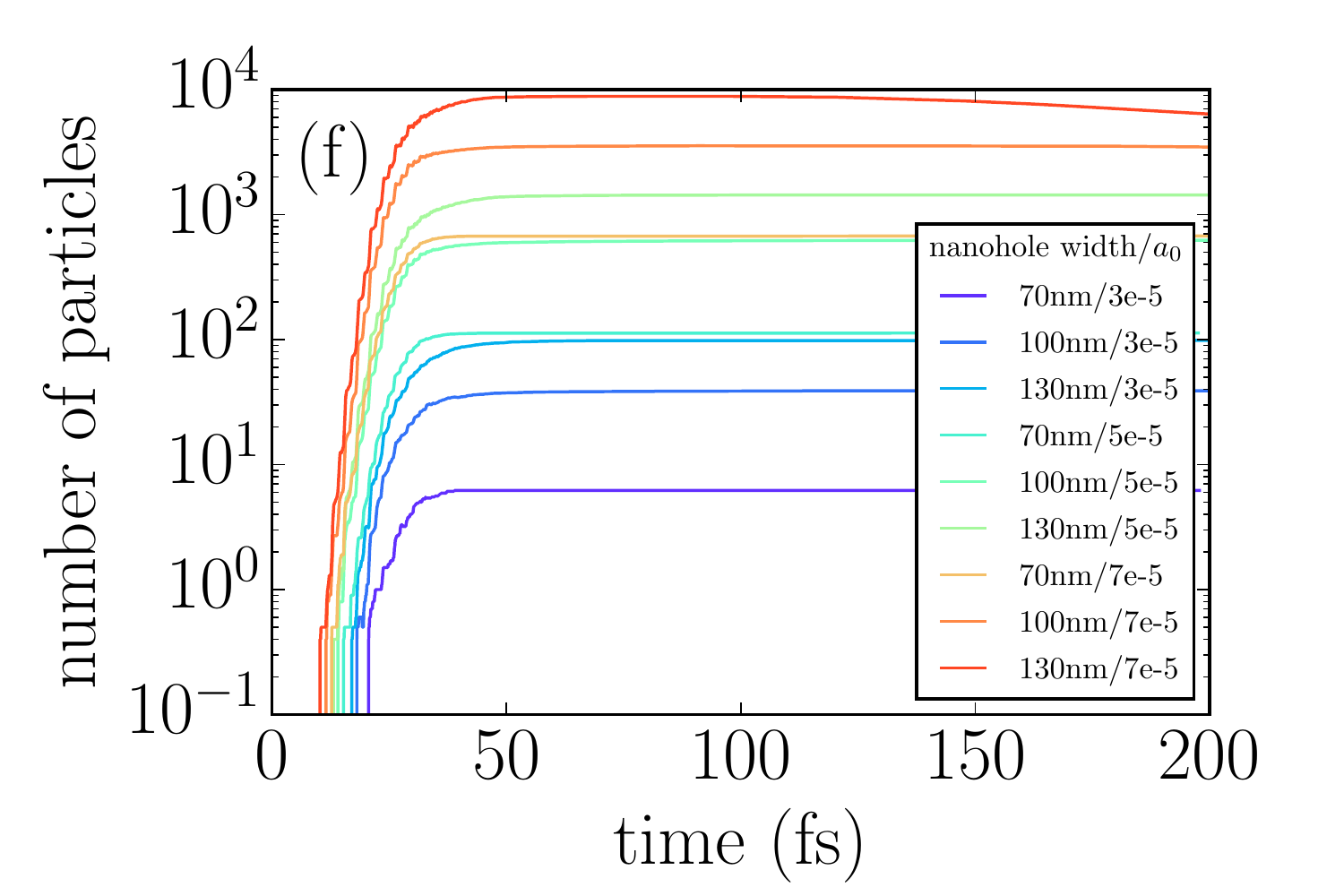}
\caption{\label{fig:plot_d_w} Transverse emittance (a,b), number of electrons (c,d) recorded 100-nm from the cathode surface and for $a_0=5\times 10^{-5}$ (a, b), and time evolution of the emitted number of electrons (e,f).  Plots (a,c,e) [resp. (b,d,f)] correspond to the case of a varying nanohole depth (resp. width) with fixed width $w=155$~nm (resp. depth $d=264$~nm).}
\end{figure}

A second study consisted in exploring the dependence of the emitted-beam properties as function of the nanohole depth $d$ and rms width $w$; see Fig.~\ref{fig:plot_d_w}. The dependence on the hole depth [Fig.~\ref{fig:plot_d_w}(left column)] confirms deeper hole enable the extraction of higher charge due to the sharper edge which provide further enhancement of the E fields. Likewise, wider holes support larger emission current due to the increased emitting area [Fig.~\ref{fig:plot_d_w}(right column)] . Both parameters can be optimized to achieve the highest possible charge yield. However, as the number of particles increases, the space-charge effects aforementioned manifest themselves and eventually limit the electron emission; see Fig.~\ref{fig:plot_d_w}(e,f). Finally, we find that the transverse emittance generally increases with the nanohole depth and width; see Fig. \ref{fig:plot_d_w}(a,b). 
\begin{figure}[hhhhhh!!!!]
	\includegraphics[width=0.49\linewidth]{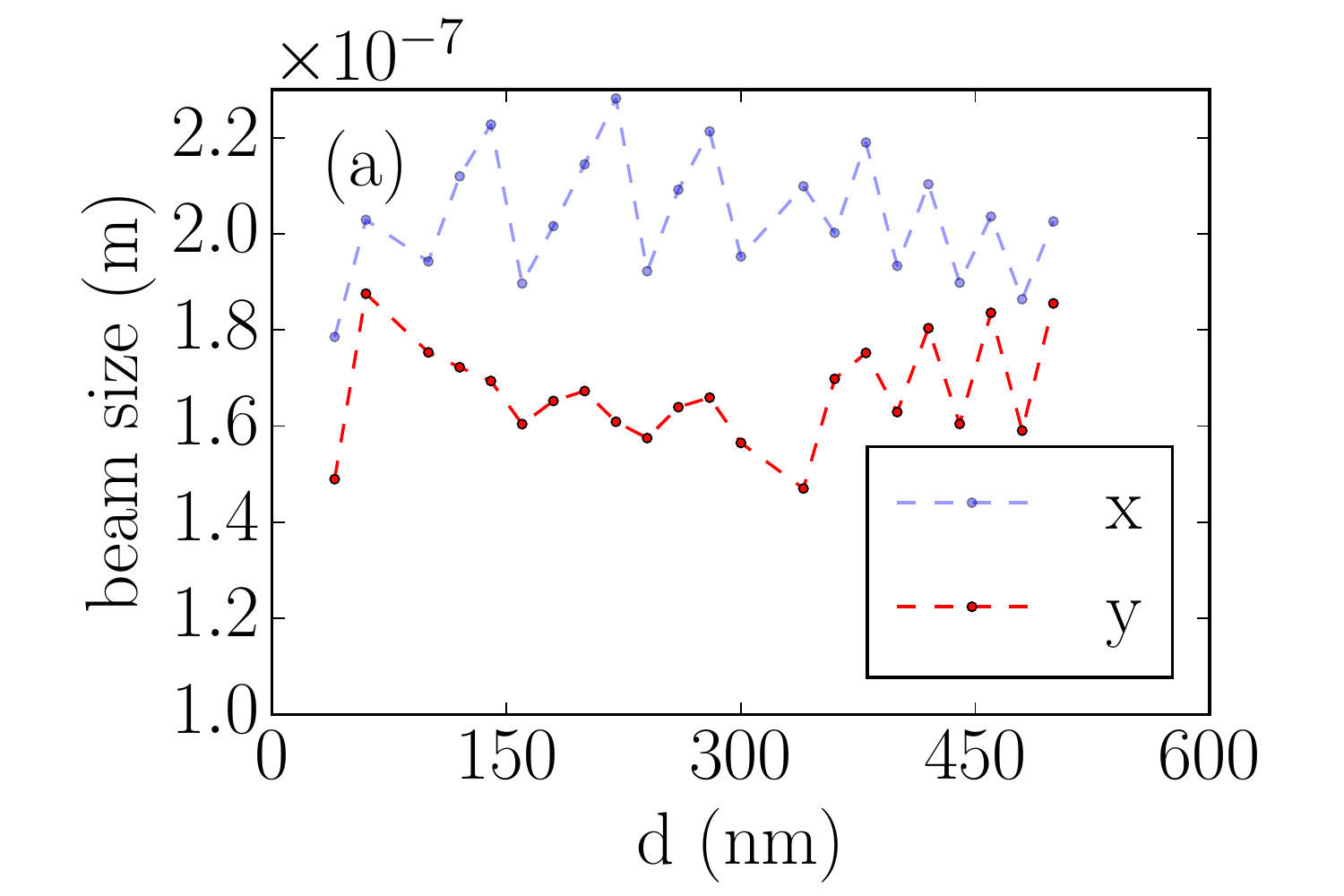}
	\includegraphics[width=0.49\linewidth]{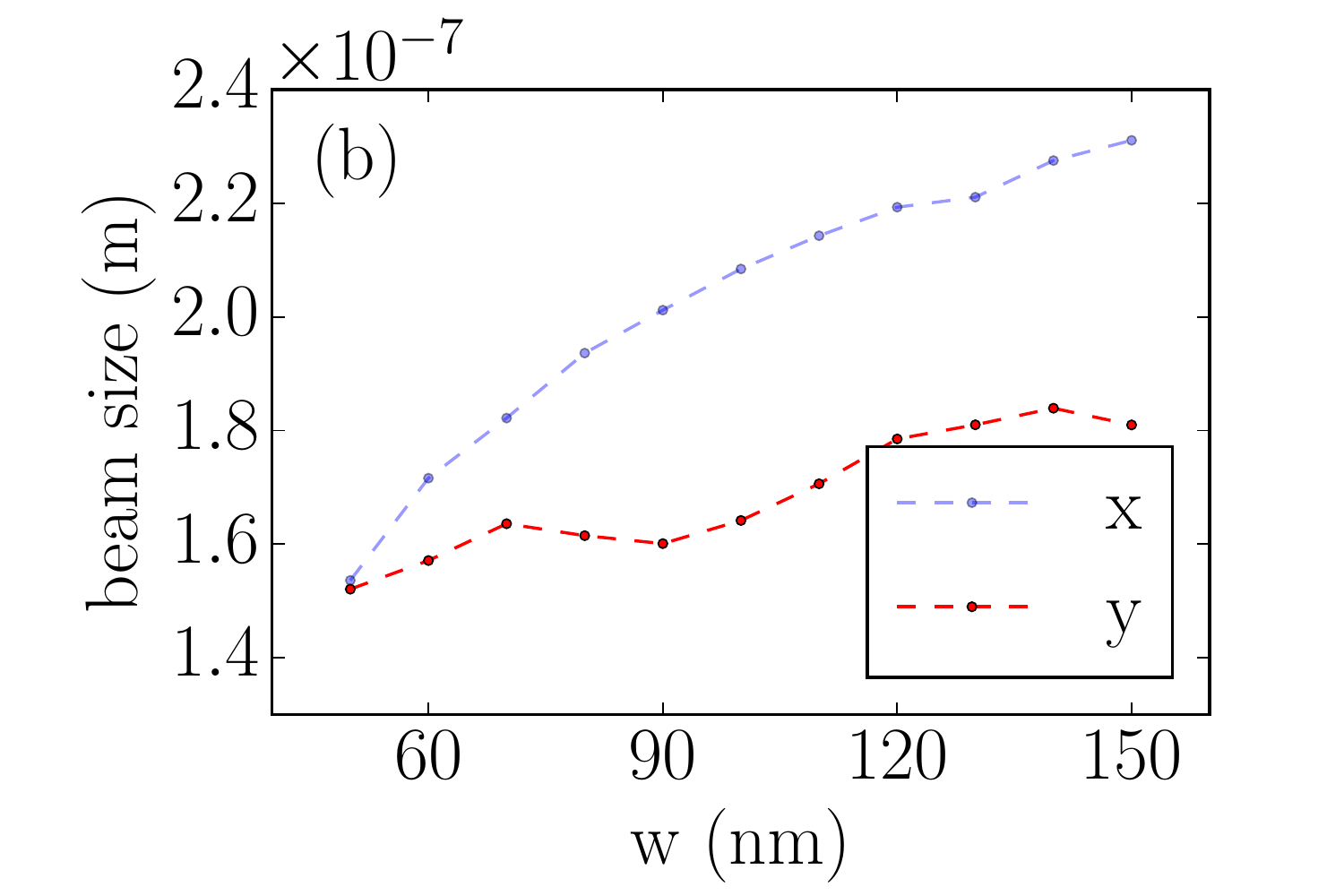}
	\includegraphics[width=0.49\linewidth]{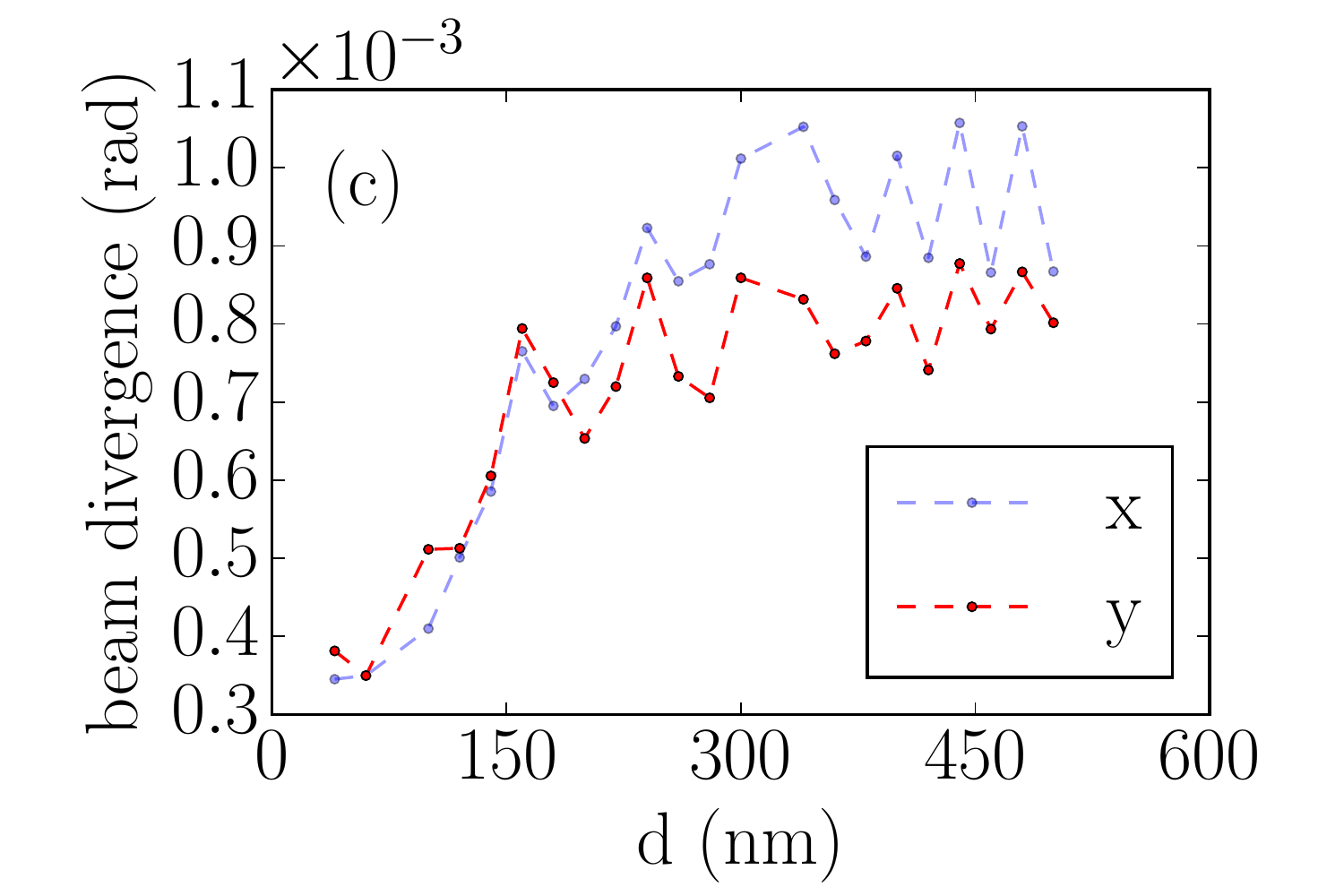}
	\includegraphics[width=0.49\linewidth]{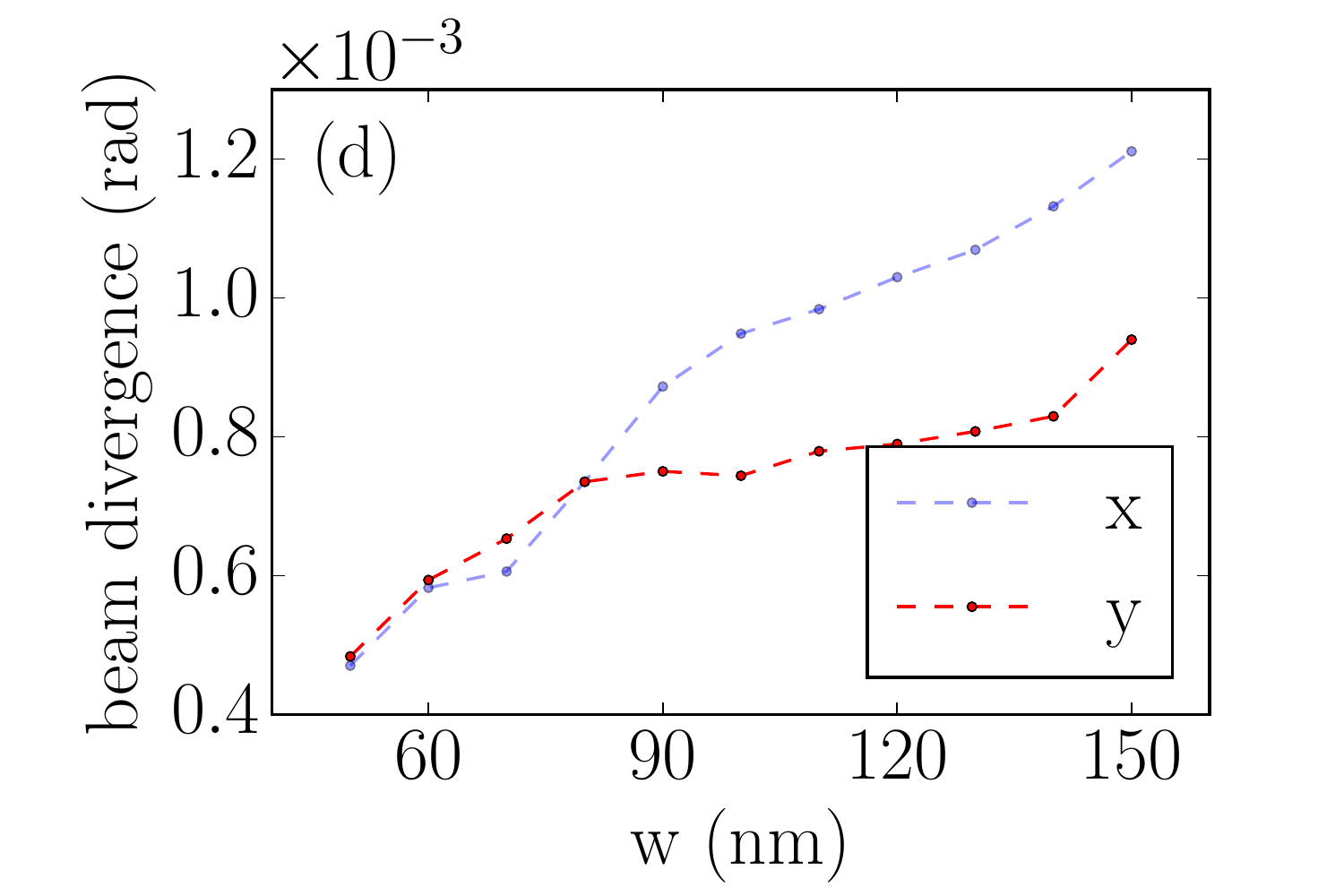}
	\caption{\label{fig:plot_d_w_2} Beam size (a,b) and divergence (c,d) recorded 100-nm from the cathode surface and for $a_0=5\times 10^{-5}$ (a, b).  Plots (a,c) [resp. (b,d)] correspond to the case of a varying nanohole depth (resp. width) with fixed width $w=155$~nm (resp. depth $d=264$~nm).}
\end{figure}
To shed further light on the emittance evolution, Figure~\ref{fig:plot_d_w_2} summarizes the evolution of beam size and divergence as function of the nanohole parameters. Increasing the depth of the hole mainly affect the divergence as electron are emitted over a surface with larger curvature; see Fig. \ref{fig:plot_d_w_2}(a,c). In contrast an increase of the width leads to larger beam sizes and divergences as summarized in Fig. \ref{fig:plot_d_w_2}(b,d).

\section{Acceleration in an RF gun}

To simulate the beam property of the entire cathode (e.g. an array or regularly spaced nanohole), the particle ensemble simulated for the single-nanohole model is replicated over the array.  This new distribution is then used as a starting distribution in an RF gun and the corresponding dynamics simulated with the fast PIC program~\astra~\cite{astra}. To date our efforts have focused on transversely imaging the cathode-array structure downstream of an accelerator beamline with $4\times4$ transfer matrix $R$. 
Given the Courant-Snyder parameters [for the horizontal plane ($\alpha_{x,1}$, $\beta_{x,1}$)] associated to a single beamlet (formed from a nanohole) along with the parameters computed of the entire beam ($\alpha_{x}$, $\beta_{x}$) we found a general relationship that ensure single-particle imaging in the horizontal plane to be~\cite{Anusorn} $R_{11}/R_{12}=\alpha_{x,1}/\beta_{x,1}$ and $R_{21}/R_{22}=\alpha_{x}/\beta_{x}$. It should be noted that the beamlet and full-beam parameters are actually connected via the array geometry and single-beamlet emittance~\cite{Rhee}. The latter set of equations respectively force $(i)$ each beamlet to be at a waist (we assume all the beamlets have identical parameters) and $(ii)$ the entire beam to be collimated. A similar equation can be written for the vertical plane. 
\begin{figure}[ttt!!!!!!]
\begin{center}
	\includegraphics[width=0.49\linewidth]{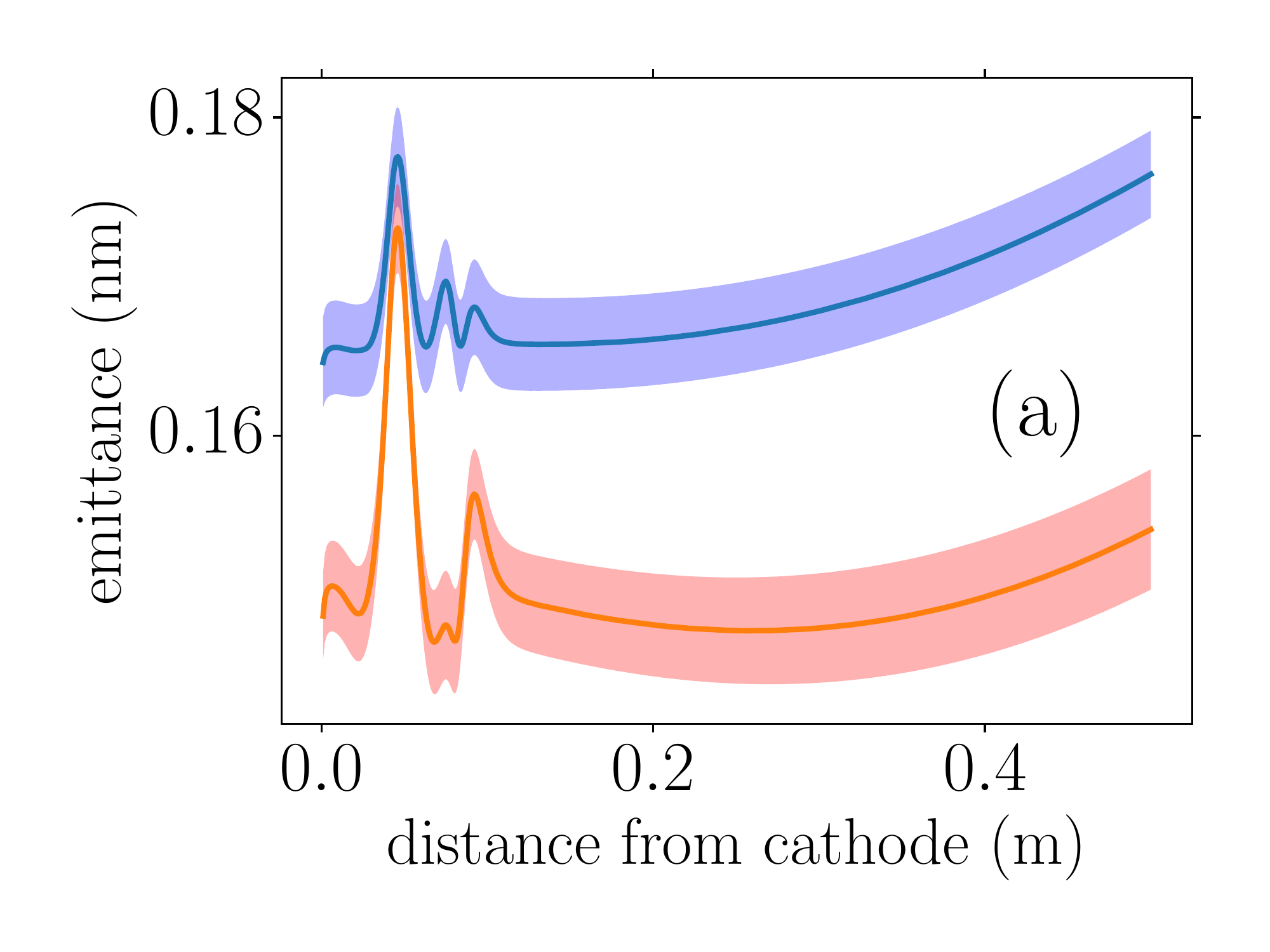}
	\includegraphics[width=0.49\linewidth]{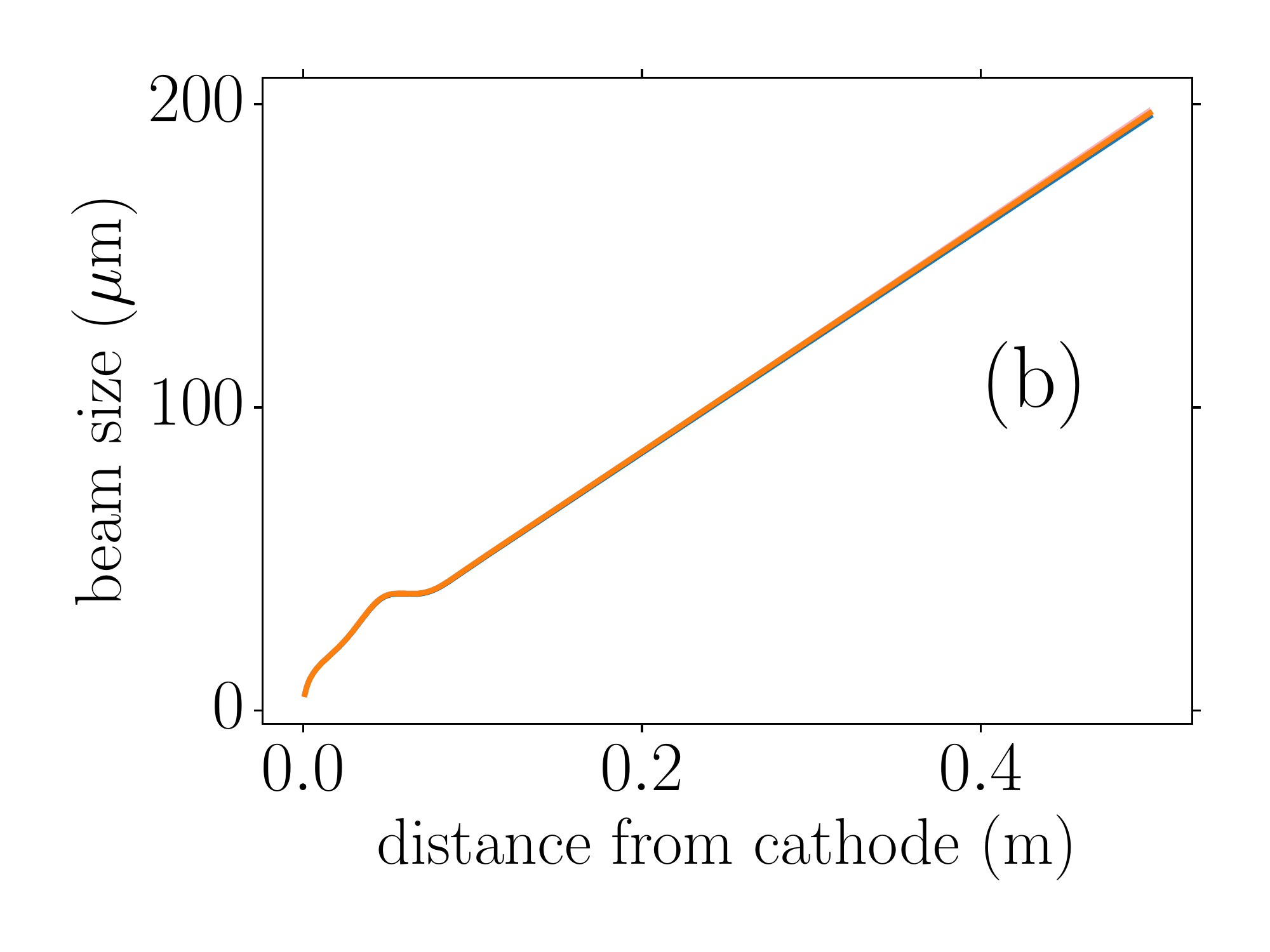}
	\includegraphics[width=0.49\linewidth]{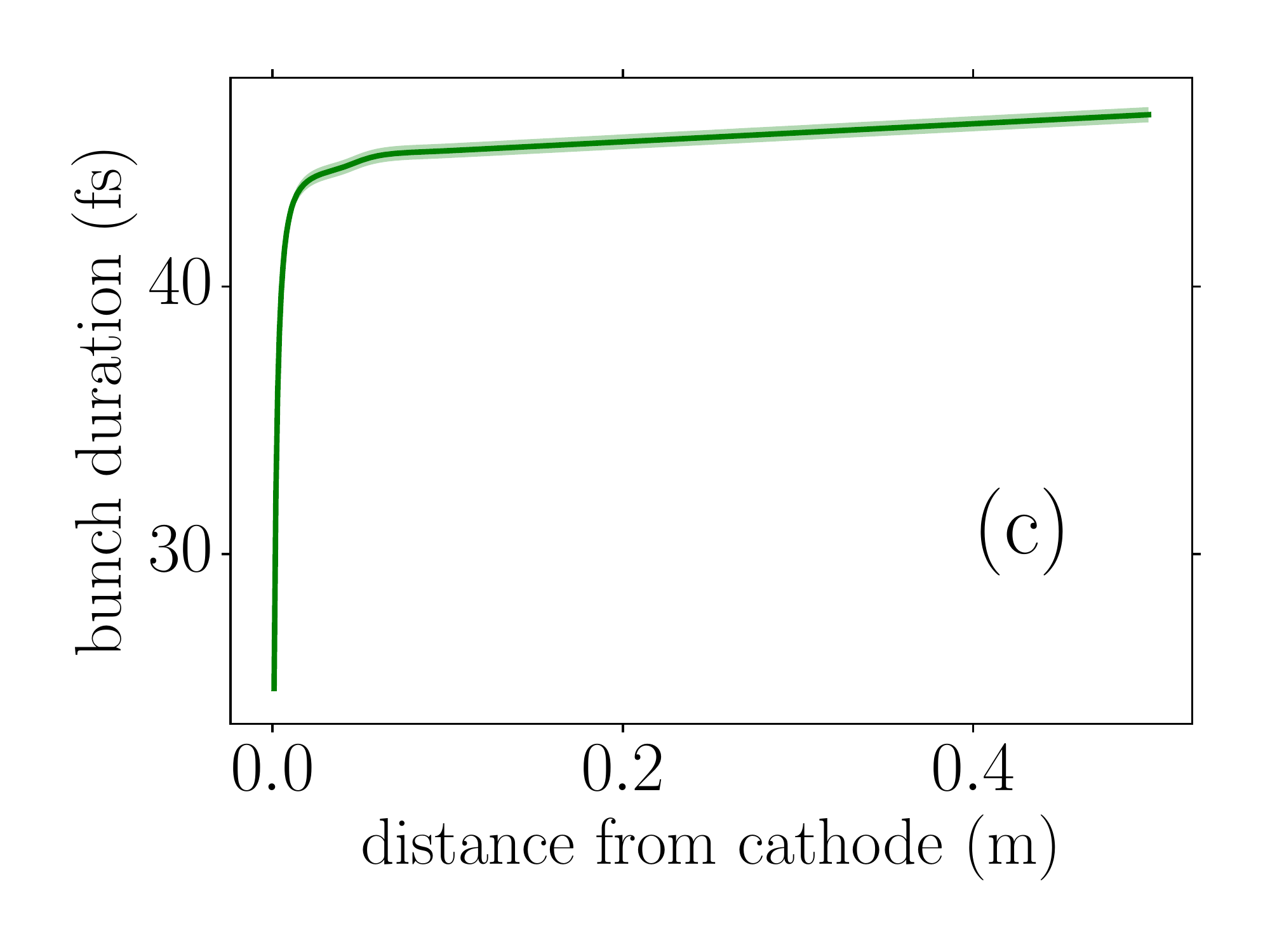}
	\includegraphics[width=0.49\linewidth]{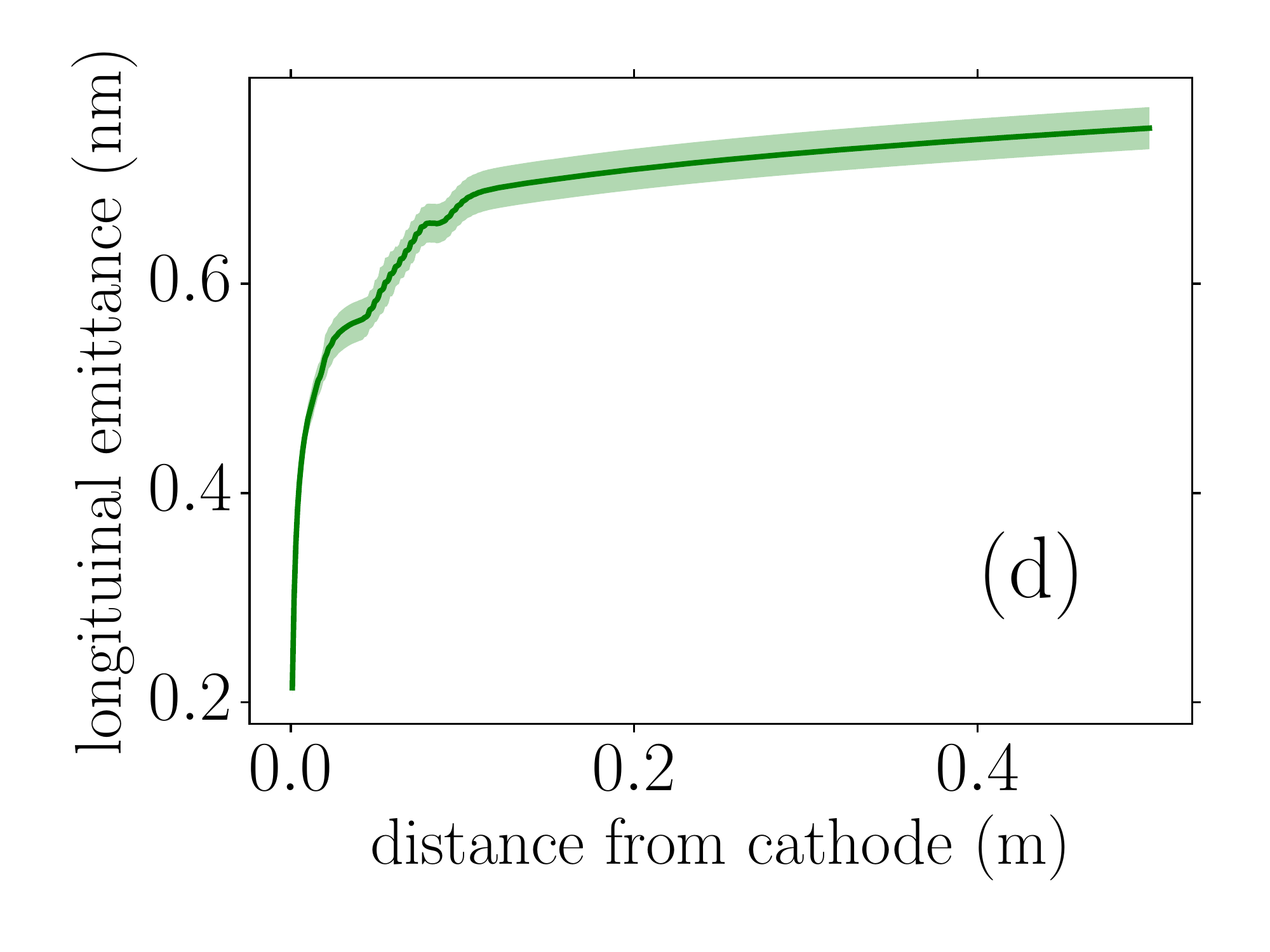}
\end{center}
\caption{\label{fig:gunSimu} Evolutions of the transverse emittances (a), beam sizes (b), bunch duration (c) and longitudinal emittance (d)    along the beamline as the beam formed by one nanohole is accelerated in a RF gun (the gun E-field extends up to $\sim0.15$~m). The beam parameters were evaluated for several random realizations of the microparticle distribution shown in Fig.~\ref{fig:fieldenhancement}(left). The solid trace and shaded areas respectively correspond to the beam value and rms spread of the evaluated parameters. }
\end{figure}
\begin{figure}[hhhhhh!!!!!!]
\begin{center}
\includegraphics[width=0.49\linewidth]{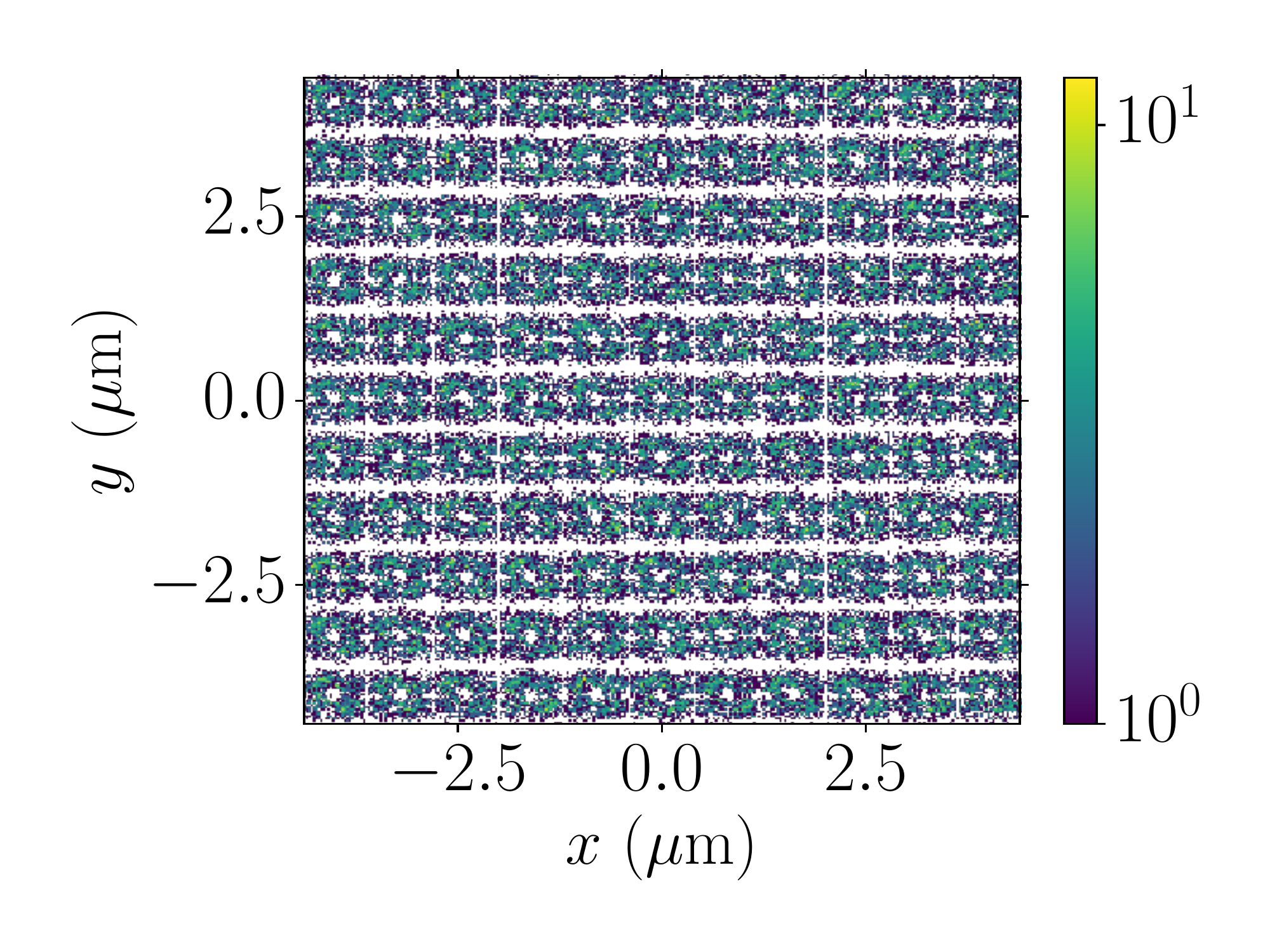}
\includegraphics[width=0.49\linewidth]{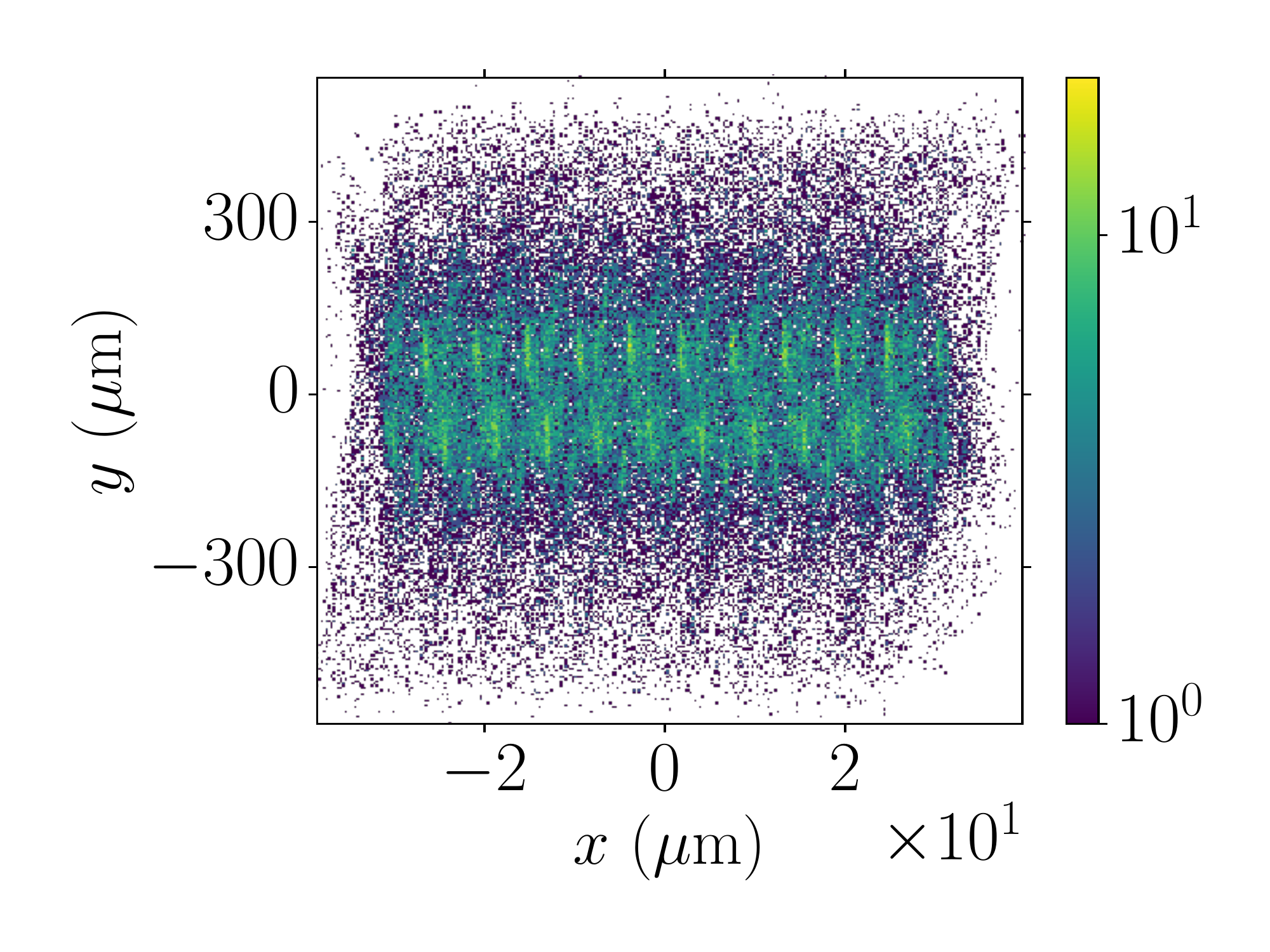}
\end{center}
\caption{\label{fig:array} Transverse initial distribution at 100-nm from a $11\times 11$ nanohole array (left) and distribution produced after 3 quadrupole magnets (right) set to satisfy the required matching conditions for horizontal imaging (space charge force are turned off in the RF gun).}
\end{figure}

In order to verify our approach, we consider a conventional LCLS-type S-band RF gun ($f=2.856$~GHz) and take the beam initial distribution to follow the case displayed in Fig.~\ref{fig:fieldenhancement}(left). As a first step we investigate the evolution of the distribution formed from one nanohole along the beamline; see Fig~\ref{fig:gunSimu}. To generate the initial distribution, the ``microparticle" distribution (each particle in the simulation represents 1/10th of an electron) generated by {\sc warp} is randomly sampled to extract $\sim 10^3$ electrons (corresponding to a charge of $\sim 160$~aC) and 10 random realizations of the distribution are considered. The beam parameters produced by one nanohole are maintained during acceleration to relativistic energies through the RF gun (final kinetic energy $K=5.6$~MeV). 

Finally we numerically confirmed the validity of the matching conditions for horizontally imaging a $11\times 11$ array of nanoholes; see  Fig.~\ref{fig:array}. As expected the resulting distribution after a beamline composed of three quadrupole magnets consists of horizontal stripes; Fig.~\ref{fig:array}(right). It should be noted that for these simulations space charge was turned off. Subsequent investigation including space-charge effects indicates a significant degradation of the imaging. We are presently addressing ways of mitigating these degradations. 

\section{Acknowledgments}
We would like to express our gratitude to Drs. D. Grote and J.-L Vay of BerkeleyLab for their help with the \warp  program. This material is based upon work supported by the US Department of Energy (DOE) under contract DE-SC0009656 with Radiabeam Technologies and by the National Science Foundation under Grant PHY-1535401 with Northern Illinois University.

\end{document}